\documentclass[conference]{IEEEtran}
\IEEEoverridecommandlockouts
\usepackage{cite}
\usepackage{amsmath,amssymb,amsfonts}
\usepackage{algorithmic}
\usepackage{graphicx}
\usepackage{textcomp}
\usepackage{xcolor}
\usepackage{hyperref}
\usepackage{multirow}
\usepackage{subcaption}
\def\BibTeX{{\rm B\kern-.05em{\sc i\kern-.025em b}\kern-.08em
    T\kern-.1667em\lower.7ex\hbox{E}\kern-.125emX}}
\begin{document}

\title{Season-Independent PV Disaggregation Using Multi-Scale Net Load Temporal Feature Extraction and Weather Factor Fusion\\
% {\footnotesize \textsuperscript{*}Note: Sub-titles are not captured for https://ieeexplore.ieee.org  and
\thanks{*Corresponding author: Hao Wang (hao.wang2@monash.edu).}
\thanks{This work was supported in part by the Australian Research Council (ARC) Discovery Early Career Researcher Award (DECRA) under Grant DE230100046.}
}

\author{\IEEEauthorblockN{1\textsuperscript{st} {Xiaolu Chen}
\IEEEauthorblockA{\textit{School of Computer Science and Technology} \\
\textit{University of Electronic Science and Technology of China}\\
Chengdu, China \\
jzzcqebd@gmail.com}
\and
\IEEEauthorblockN{2\textsuperscript{nd} Chenghao Huang}
\IEEEauthorblockA{\textit{Department of Data Science and AI}\\ \textit{Faculty of Information Technology} \\
\textit{Monash University}\\
Melbourne, Australia \\
Chenghao.Huang@monash.edu}}
\and
\IEEEauthorblockN{3\textsuperscript{rd} Yanru Zhang}
\IEEEauthorblockA{\textit{School of Computer Science and Technology} \\
\textit{University of Electronic Science and Technology of China}\\
Chengdu, China \\
\textit{Shenzhen Institute for Advanced Study}\\
Shenzhen, China \\
yanruzhang@uestc.edu.cn}
\and
\IEEEauthorblockN{4\textsuperscript{th} Hao Wang*}
\IEEEauthorblockA{\textit{Department of Data Science and AI}\\ \textit{Faculty of Information Technology} \\
\textit{and Monash Energy Institute} \\
\textit{Monash University}\\
Melbourne, Australia \\
hao.wang2@monash.edu}
}

\maketitle

\begin{abstract}
With the advancement of energy Internet and energy system integration, the increasing adoption of distributed photovoltaic (PV) systems presents new challenges on smart monitoring and measurement for utility companies, particularly in separating PV generation from net electricity load. Existing methods struggle with feature extraction from net load and capturing the relevance between weather factors. 
This paper proposes a PV disaggregation method that integrates Hierarchical Interpolation (HI) and multi-head self-attention mechanisms. By using HI to extract net load features and multi-head self-attention to capture the complex dependencies between weather factors, the method achieves precise PV generation predictions. Simulation experiments demonstrate the effectiveness of the proposed method in real-world data, supporting improved monitoring and management of distributed energy systems.
\end{abstract}


\section{Introduction}

% Background
With the increasing adoption of distributed solar photovoltaic (PV) systems, an expanding number of residential prosumers, who both produce and consume electricity, are generating electricity through their PV installations. A prevalent challenge faced by utility companies is that many of these prosumers are equipped with smart meters that measure only the net electricity load, i.e. the difference between actual electricity consumption and generation, without providing separate records of PV generation~\cite{8960431}. This lack of detailed measurement complicates efforts to assess the impact of distributed generation on the power grid~\cite{7784836} and to maintain supply-demand balance~\cite{9159864}.

PV disaggregation is the process of separating PV generation from net electricity load to estimate the amount of solar electricity produced by prosumers equipped with PV systems~\cite{ERDENER2022112224}. Precise PV disaggregation is essential for utility companies to monitor distributed generation effectively. Moreover, it is critical in enabling effective demand response strategies and supporting the continued expansion of distributed renewable energy systems.

A review of PV disaggregation literature reveals several approaches and challenges. Model-based methods, which rely on physical models assuming prior knowledge of PV system configurations, were initially common. However, they struggled with real-world inconsistencies, particularly in Behind-The-Meter (BTM) setups~\cite{9360481,9409943}. As a result, data-driven methods have gained prominence, avoiding explicit physical models and offering greater flexibility.
Among these, machine learning frameworks are frequently used. Saeedi et al. proposed a model using weather data and minimal PV sensors to estimate PV generation, focusing on time dependency and dimensionality reduction~\cite{9360481}. Bu et al. used Gaussian mixture models with low-resolution smart meter data to decompose monthly solar estimates into hourly profiles~\cite{9409943}. Neighbor-based approaches like Chen et al.'s method use neighboring users' electricity data as proxies, without needing weather data or specific model assumptions~\cite{9735454}.

Another significant trend involves probabilistic and real-time estimation methods. Yi et al. introduced a Bayesian dictionary learning model that provides uncertainty measures and uses real-time calculations for reliability~\cite{9477124}. Pan et al. developed an unsupervised learning method that refines PV disaggregation using meteorological data~\cite{PAN2022118450}. Wang et al. presented a context-supervised source separation technique to estimate PV generation and battery behavior~\cite{9684963}.
Finally, hybrid and event-based approaches have also been explored. Liu et al. combined dictionary learning with event-based detection to improve non-intrusive disaggregation accuracy~\cite{LIU2022107887}. The work of~\cite{9766031} used LSTM networks to refine PV generation and consumption models based on net load and weather data. Stratman et al. proposed a two-stage framework for net load forecasting with limited observability, using a compensator to adjust the forecast~\cite{10125003}.

Although existing research has made progress in PV disaggregation, several challenges remain in effectively separating PV generation from net load data. The two main challenges posed by previous works on PV disaggregation are summarized as follows.

\begin{itemize}
    \item \textbf{Limited feature extraction from net electricity load:} Traditional methods often struggle to effectively extract features from net load data, which leads to suboptimal disaggregation performance.
    \item \textbf{Insufficient capture of relevance between weather factors:} Insufficient capture of relevance between weather factors poses a significant challenge in PV disaggregation, as PV generation is highly influenced by dynamic weather conditions. Traditional models often fail to effectively take advantage of the complex relationships between factors like irradiance, leading to less precise predictions.
\end{itemize}

% method
The goal of this work is to develop an effective PV disaggregation method that precisely predicts daily PV generation, enabling utility companies to better monitor distributed energy systems. To tackle the challenges of separately processing net electricity load and weather data, our approach integrates Hierarchical Interpolation (HI)~\cite{challu2023nhits} and multi-head self-attention mechanisms~\cite{vaswani2017attention} into a unified framework. HI extracts essential features from the net load, while the multi-head self-attention network captures relevance between various weather factors. By fusing these embeddings, the proposed method provides a robust and precise prediction of PV generation.

% contribution
The main contributions of our work can be summarized as follows.
\begin{itemize}
    \item \textbf{Hierarchical interpolation for net load feature extraction:} Our approach employs HI to process net electricity load data, transforming it into a feature embedding. This method enables precise extraction of load characteristics, which enhances the performance of the disaggregation process.
    \item \textbf{Multi-head self-attention for weather data integration:} We utilize a multi-head self-attention mechanism to capture complex relevance between various weather factors. This network effectively generates an embedding that reflects the intricate relationships among weather variables, improving the model's ability and robustness to take advantage of weather-related influences on PV generation.
    \item \textbf{Validation through real-world data simulation:} We conduct extensive simulation experiments using real-world data, demonstrating the effectiveness and robustness of our proposed method in precisely disaggregating PV generation from net load.
\end{itemize}

\section{Problem Statement}

We denote the net electricity load of a distributed solar prosumer at time $t$ as $l_t$, the corresponding PV generation as $g_t$, and the corresponding actual electricity consumption as $u_t$ which may be composed by the energy from both the grid and the PV generation. The net electricity load $l_t$ can be expressed as:
\begin{align}
    l_t = u_t-g_t.
\end{align}

Though it is important for utility companies to maintain grid safety through distributed energy monitoring, they nowadays struggle with comprehensively accessing data from all of their customers, because some customers install PV systems for self-generation, while their smart meters only record net electricity load. To monitor distributed generation, utility companies intend to estimate daily PV generation and actual electricity consumption of prosumers based on their net electricity load and corresponding weather data, since PV generation is highly related to weather issues.
Suppose utility companies aim to disaggregate PV generation across an entire day $d$ divided into $T$ time slots, each of the variables is extended into a time series over a given horizon denoted as $T$: $W_d = \{w_{d,t}\}_{t=1}^T$ for weather data, $L_d=\{l_{d,t}\}_{t=1}^T$ for net electricity load, and $G_d=\{g_{d,t}\}_{t=1}^T$ for PV generation.
The objective is to train a PV disaggregation model $f$, where the input $X_d=[W_d, L_d]$ consists of the series of weather data and net load at day $d$, and the output is the predicted series of PV generation $\hat{G}_d$, formulated as follows:
\begin{align}
    \hat{G}_d = f(W_d, L_d).
\end{align}

Generally, there are two types of prosumers, distinguished by their smart meter recordings. For Type 1 prosumers, smart meters measure both the net electricity load $l_{d,t}$ and the PV generation $g_{d,t}$. For Type 2 prosumers, smart meters only measure the net electricity load $l_{d,t}$.
Let $P_1$ and $P_2$ represent the sets of the user IDs of Type 1 and Type 2 prosumers, respectively.
By training the model $f$ using the dataset $\mathcal{D}_1$ consisting of Type 1 prosumers in total $D$ days, $\mathcal{D}_1=\{X_d^i, G_d^i\}_d^D, i \in P_1$, where $G_d^i$ is treated as the ground truth for supervised learning, utility companies can predict the PV generation using the input data from the dataset $\mathcal{D}_2$ consisting of Type 2 prosumers, $\mathcal{D}_2=\{X_d^j\}_d^D, j \in P_2$.

\section{Methodology}\label{Sec: method}

\begin{figure}
    \centering
    \includegraphics[width=0.99\linewidth]{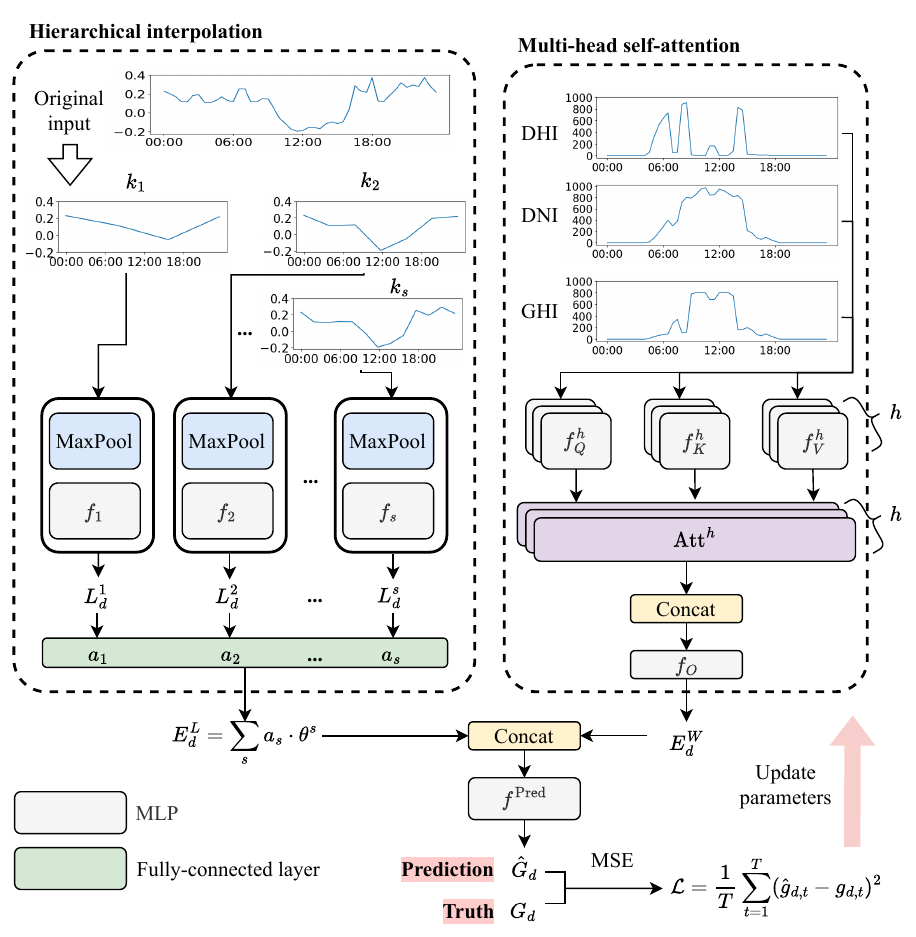}
    \caption{The illustration of the proposed method.}
    \label{fig:method}
\end{figure}

In this section, we propose a DL-based approach for PV disaggregation, which combines hierarchical interpolation (HI) and multi-head self-attention mechanisms to process both the net electricity load and weather data. Briefly, HI is conducted to separately process net electricity load into a feature embedding. Furthermore, a multi-head self-attention feed-forward network is constructed to effectively capture interconnection between multiple weather factors and output another embedding. Finally, the embeddings from both sources are fused to generate an precise prediction of the daily PV generation, which can help utility companies better monitor distributed energy systems. The illustration of the proposed method is shown in Fig.~\ref{fig:method}.

\subsection{HI for Net Electricity Load}

To capture the temporal patterns in the net electricity load, we employ the HI technique. The net electricity load $L_d$ of day $d$ is decomposed into multiple scales, where each scale captures a different temporal pattern. This is achieved through a multi-rate signal decomposition process followed by interpolation at each scale. 

At scale $s$, the load data $L_D$ is subsampled as follows:
\begin{align}
    L^s_d = \text{MaxPool}(L_d, k_s),
\end{align}
where $k_s$ is the kernel size controlling the subsampling rate. Larger $k_s$ captures long-term temporal dependency, while smaller $k_s$ captures short-term one.

Then, the interpolated load at each scale is generated using a Multi-Layer Perceptron (MLP) $f_s$, which outputs forward and backward interpolation coefficients:
\begin{align}
    \theta^{s} = f_s(L_d^{s}).
\end{align}
The final embedding $E_L$ for the net electricity load is obtained by combining the interpolation coefficients across multiple scales:
\begin{align}
    E^L_d = \sum_s a_s \cdot \theta^s,
\end{align}
where $a_s$ is a learnable weight associated with scale $s$. This embedding $E^L_d$ captures the temporal structure of the load data over multiple scales.

\subsection{Multi-Head Self-Attention for Weather Data}

The multi-head self-attention mechanism is a key component in transformer architectures, designed to capture complex dependencies in data sequences by focusing on different parts of the input simultaneously. Self-attention calculates attention scores to determine the importance of each part of the input sequence relative to others.

The weather data in our study consists of three specific components, including Direct Normal Irradiance (DNI) $W^{\text{DNI}}_d = \{w^{\text{DNI}}_{d,t}\}_{t=1}^{T}$, Diffuse Horizontal Irradiance (DHI)  $W^{\text{DHI}}_d = \{w^{\text{DHI}}_{d,t}\}_{t=1}^{T}$, and Global Horizontal Irradiance (GHI) $W^{\text{GHI}}_d = \{w^{\text{GHI}}_{d,t}\}_{t=1}^{T}$. The three weather time series are processed using a multi-head self-attention mechanism. Each of these time series is treated as an input sequence, and the self-attention mechanism is used to capture the dependencies between these weather components.

The input features are transformed into three vectors.
Query $Q$ represents what each position is trying to find in the sequence.
Key $K$ represents characteristics of each position that others can relate to.
Value $V$ contains the actual information to be aggregated based on the attention scores. 
The attention score is calculated as the dot product of the query and key vectors, which is then normalized using a softmax function.
Suppose $H$ heads are adopted in the multi-head self-attention network. For each head $h$, the inputs of day $d$ $W^{\text{DNI}}_d, W^{\text{DHI}}_d, W^{\text{GHI}}_d$ are first projected into query $Q$, key $K$, and value $V$ representations using MLPs:
\begin{align}
Q^h = f_Q^h(W^{\text{DNI}}_d, W^{\text{DHI}}_d, W^{\text{GHI}}_d),\\
K^h = f_K^h(W^{\text{DNI}}_d, W^{\text{DHI}}_d, W^{\text{GHI}}_d),\\
V^h = f_V^h(W^{\text{DNI}}_d, W^{\text{DHI}}_d, W^{\text{GHI}}_d).
\end{align}

Then, the attention weights for each head $h$ are calculated by taking the scaled dot product of the query $Q$ and key $K$:
\begin{align}
    \text{Att}^h = \text{Softmax}\left( \frac{Q^h K^{h\top}}{\sqrt{d_h}} \right) V^h,
\end{align}
where $d_h$ is the dimensionality of each head. The softmax function ensures that the attention weights sum to one across all time steps.

In multi-head self-attention, multiple independent heads perform self-attention in parallel. Each head has its own set of $Q$, $K$, and $V$ matrices, allowing the mechanism to capture diverse patterns within the input. By focusing on different aspects of the data in each head, the model can learn various relationships and dependencies. The outputs from all heads are concatenated and transformed through a linear projection.

Next, the outputs from multiple attention heads are concatenated to form the final weather embedding $E_{W,d}$:
\begin{align}
    E^W_d = f_O\big(\text{Concat}(\text{Att}^1, \text{Att}^2, \dots, \text{Att}^H)\big),
\end{align}
where $f_O$ is an MLP. This embedding $E^W_d$ represents the combined features of all weather components and their interactions.

\subsection{Embedding Fusion and Final Prediction}

The embeddings from the net electricity load $E^L_d$ and weather data $E^W_d$ of day $d$ are concatenated to form a joint representation:
\begin{align}
    E^{\text{Fuse}} = \text{Concat}(E^L_d, E^W_d).
\end{align}
This concatenated embedding $E^{\text{Fuse}}$ is then passed through a fully connected layer to predict the disaggregated PV generation over the day $d$:
\begin{align}
    \hat{G}_d = (\hat{g}_{d,1},...,\hat{g}_{d,T}) = \sigma[f^{\text{Pred}}(E^{\text{Fuse}})],
\end{align}
where $\sigma$ is an activation function, e.g., ReLU.

To train our model for PV disaggregation, we use the Mean Squared Error (MSE) as the loss function, calculated as:
\begin{align}
    \mathcal{L} = \frac{1}{T}\sum^T_{t=1}(\hat{g}_{d,t}-g_{d,t})^2.
\end{align}
The goal is to minimize the difference between the predicted PV generation and the actual PV generation over the $T$ time slots in day $d$.

\section{Experiments}

In this section, we present the datasets and evaluation metrics used, followed by a detailed description of the benchmark methods selected for our experiments.  We then analyze and discuss the performance of the benchmark methods and the proposed method, including comparisons across different evaluation metrics.

\subsection{Dataset and Metrics}

The dataset used for the experiments is the Solar Home Electricity Data~\cite{ratnam2017residential} provided by Ausgrid’s electricity network, which contains three years of half-hourly electricity data for 300 randomly selected solar homes, covering the period from July 1, 2010, to June 30, 2013.
To ensure data quality, homes with extremely high or low electricity consumption or PV generation during the first year are excluded. The dataset includes three key consumption categories: general consumption, representing household electricity use excluding PV generation and controlled loads; controlled load consumption, which reflects electricity usage under off-peak tariffs; and gross generation, which records the total electricity generated by the solar system, separate from household consumption. The weather data for the 300 solar homes are obtained from the National Solar Radiation Database~\cite{NREL}, a continuous and comprehensive dataset with high temporal (hourly or half-hourly) and spatial resolution, covering a wide range of meteorological data. In this experiment, three key solar radiation measurements are used: global horizontal irradiance, direct normal irradiance, and diffuse horizontal irradiance.
These datasets are preprocessed using Z-score normalization.

In this experiment, we select two evaluation metrics, Mean Absolute Error (MAE) and Root Mean Square Error (RMSE), to assess the accuracy of the model's output, which is a time series consisting of 48 data points. MAE measures the average magnitude of errors between predicted and actual values, providing an intuitive sense of overall prediction accuracy. It is defined as:
\begin{align}
    \text{MAE} = \frac{1}{T}\sum_{t=1}^T |\hat{y}_t-y_t|,
\end{align}
where $T=48$ is the total number of time points, $\hat{y}_t$ is the predicted value at time $t$, and $y_t$ is the ground-truth value at time $t$. RMSE, on the other hand, emphasizes larger deviations by squaring the differences between predicted and actual values before averaging and taking the square root. It is defined as:
\begin{align}
    \text{RMSE} =\sqrt{ \frac{1}{T}\sum_{t=1}^T (\hat{y}_t-y_t)^2}.
\end{align}

\subsection{Benchmarks for Comparison}

In this experiment, we evaluate several benchmark methods to assess their performance in time series forecasting. K-Nearest Neighbors (KNN)~\cite{weinberger2009distance} is a simple and effective algorithm that classifies data points based on the proximity to their neighbors. Random Forest (Forest)~\cite{breiman2001random} is an ensemble learning method that aggregates predictions from multiple decision trees to improve accuracy and handle complex data structures. InceptionTime~\cite{ismail2020inceptiontime} leverages a deep learning architecture with inception modules, known for capturing multi-scale features in time series data. LSTM-FCN~\cite{KARIM2019237} combines Long Short-Term Memory (LSTM) networks with Fully Convolutional Networks (FCN) to capture both temporal dependencies and spatial features. Multi-head Attention Convolutional Neural Network (MACNN)~\cite{CHEN2021126} integrates attention mechanisms with convolutional layers to focus on significant features while processing time series data. These diverse methods provide a comprehensive comparison of traditional and advanced techniques in forecasting and modeling time series data.

\subsection{Result Analysis}

\begin{table}[]
\centering
\caption{Comparative test results.}
\label{table:result}
\begin{tabular}{cccc}
\hline\hline
\textbf{Season}                  & \textbf{Method} & \textbf{MAE(KWh)} & \textbf{RMSE(KWh)} \\ \hline
\multirow{6}{*}{\textbf{Summer}} & KNN             & 0.0429            & 0.0971             \\
                                 & Forest          & 0.0338            & 0.0722             \\
                                 & LSTMFCN         & 0.0573            & 0.0987             \\
                                 & MACNN           & 0.0557            & 0.1001             \\
                                 & Inceptiontime   & 0.0632            & 0.1119             \\ \cline{2-4} 
                                 & Proposed        & \textbf{0.0321}   & \textbf{0.0692}    \\ \hline
\multirow{6}{*}{\textbf{Autumn}} & KNN             & 0.0360            & 0.0920             \\
                                 & Forest          & \textbf{0.0293}   & 0.0702             \\
                                 & LSTMFCN         & 0.0526            & 0.0945             \\
                                 & MACNN           & 0.0494            & 0.0921             \\
                                 & Inceptiontime   & 0.0561            & 0.1016             \\ \cline{2-4} 
                                 & Proposed        & 0.0302            & \textbf{0.0687}    \\ \hline
\multirow{6}{*}{\textbf{Winner}} & KNN             & 0.0383            & 0.0994             \\
                                 & Forest          & 0.0307            & 0.0755             \\
                                 & LSTMFCN         & 0.0581            & 0.1061             \\
                                 & MACNN           & 0.0517            & 0.0982             \\
                                 & Inceptiontime   & 0.0538            & 0.0985             \\ \cline{2-4} 
                                 & Proposed        & \textbf{0.0292}   & \textbf{0.0733}    \\ \hline
\multirow{6}{*}{\textbf{Spring}} & KNN             & 0.0409            & 0.0975             \\
                                 & Forest          & 0.0347            & 0.0745             \\
                                 & LSTMFCN         & 0.0610            & 0.1038             \\
                                 & MACNN           & 0.0556            & 0.0994             \\
                                 & Inceptiontime   & 0.0633            & 0.1122             \\ \cline{2-4} 
                                 & Proposed        & \textbf{0.0321}   & \textbf{0.0724}    \\ 
\hline\hline
\end{tabular}
\end{table}

\begin{figure}[htbp]
    \centering
    \begin{subfigure}[b]{\linewidth}
        \centering
        \includegraphics[width=0.8\linewidth]{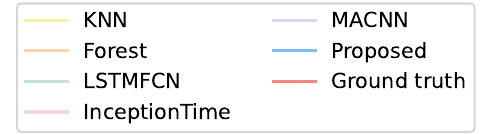}
    \end{subfigure}
    \vspace{0.05\linewidth}
    \begin{subfigure}[b]{\linewidth}
        \centering
        \includegraphics[width=0.99\linewidth]{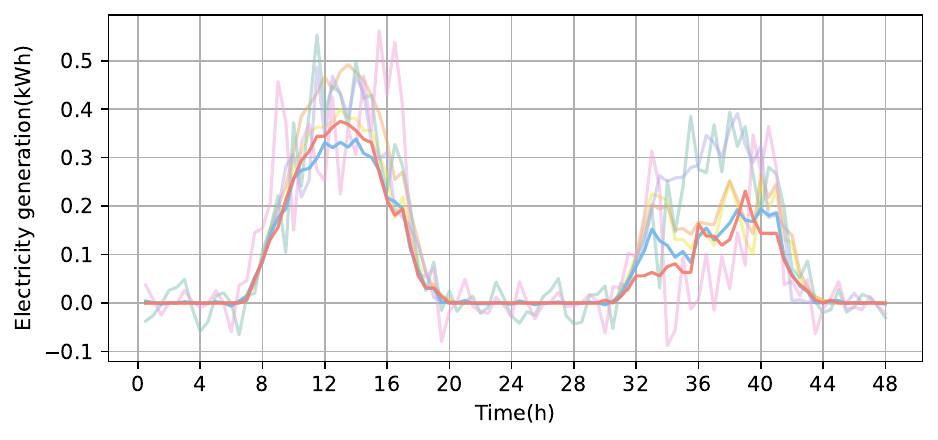}
        \caption{Summer}
    \end{subfigure}
    \vspace{0.05\linewidth}
    \begin{subfigure}[b]{\linewidth}
        \centering
        \includegraphics[width=0.99\linewidth]{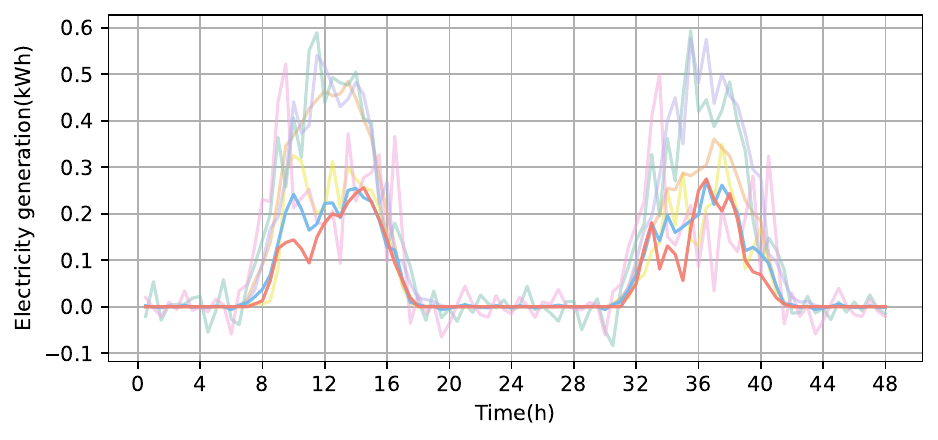}
        \caption{Autumn}
    \end{subfigure}
    \vspace{0.05\linewidth}
    \begin{subfigure}[b]{\linewidth}
        \centering
        \includegraphics[width=0.99\linewidth]{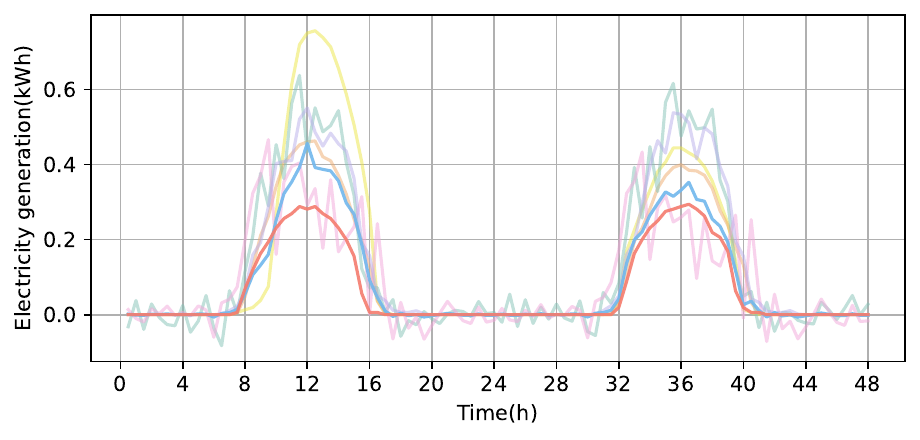}
        \caption{Winter}
    \end{subfigure}
    \vspace{0.05\linewidth}
    \begin{subfigure}[b]{\linewidth}
        \centering
        \includegraphics[width=0.99\linewidth]{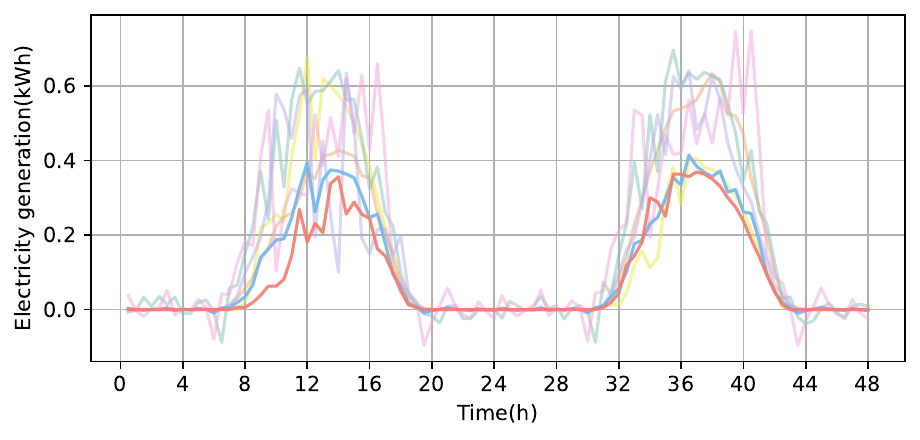}
        \caption{Spring}
    \end{subfigure}
    \caption{PV Disaggregation for Two Consecutive Days of a Specific Household}
    \label{fig:line_chart}
\end{figure}

To eliminate the effect of model performance randomness on the fair comparison, the experiments are repeated five times and the average results are presented. The results in Table~\ref{table:result} show that the proposed method performs consistently well across all seasons, demonstrating its season-independent effectiveness. For both MAE and RMSE, the proposed method achieves the lowest values in Summer, Winter, and Spring, indicating its ability to accurately disaggregate PV generation despite seasonal variations. Even in Autumn, where the Forest method slightly outperforms in MAE, the proposed method still achieves the best RMSE, highlighting its robustness in minimizing errors.

The Fig.~\ref{fig:line_chart} compares various methods for estimating PV generation across the four seasons. Each figure shows the performance of benchmark and proposed models relative to the ground truth over a 48-hour period, with data recorded at half-hour intervals. In Summer, the proposed method and KNN provide the closest approximation to the ground truth, particularly during peak PV generation periods. LSTMFCN and InceptionTime exhibit higher fluctuations and noticeable discrepancies. In Autumn, the trend is similar to that in Summer, with the proposed method aligning well with the ground truth, while LSTMFCN and MACNN frequently overestimate PV generation. In Winter, the variance between methods becomes more pronounced. The proposed method still tracks closely to the ground truth, while KNN, MACNN, and LSTMFCN show larger deviations, especially during the afternoon peak. In Spring, LSTMFCN, InceptionTime, and MACNN demonstrate significant fluctuations. Overall, the proposed method consistently exhibits robust performance across all four seasons, demonstrating its ability to estimate PV generation under varying seasonal conditions accurately. Other methods, particularly in Autumn and Winter, show greater deviations from the ground truth.

\section{Conclusion}
This paper presents an advanced PV disaggregation method that effectively addresses the challenges of separating PV generation from net load data in the context of distributed energy systems. By integrating Hierarchical Interpolation (HI) for net load feature extraction and multi-head self-attention for capturing complex weather dependencies, the proposed method achieves high precision in PV generation predictions. Simulation results using real-world data validate the robustness and effectiveness of the approach, offering utility companies a valuable tool for better monitoring and management of distributed energy systems, aligning with the growing demands of the energy Internet and energy system integration initiatives.

\bibliographystyle{IEEEtran}
\bibliography{references.bib}

\begin{thebibliography}{10}
\providecommand{\url}[1]{#1}
\csname url@samestyle\endcsname
\providecommand{\newblock}{\relax}
\providecommand{\bibinfo}[2]{#2}
\providecommand{\BIBentrySTDinterwordspacing}{\spaceskip=0pt\relax}
\providecommand{\BIBentryALTinterwordstretchfactor}{4}
\providecommand{\BIBentryALTinterwordspacing}{\spaceskip=\fontdimen2\font plus
\BIBentryALTinterwordstretchfactor\fontdimen3\font minus \fontdimen4\font\relax}
\providecommand{\BIBforeignlanguage}[2]{{%
\expandafter\ifx\csname l@#1\endcsname\relax
\typeout{** WARNING: IEEEtran.bst: No hyphenation pattern has been}%
\typeout{** loaded for the language `#1'. Using the pattern for}%
\typeout{** the default language instead.}%
\else
\language=\csname l@#1\endcsname
\fi
#2}}
\providecommand{\BIBdecl}{\relax}
\BIBdecl

\bibitem{8960431}
F.~Bu, K.~Dehghanpour, Y.~Yuan, Z.~Wang, and Y.~Zhang, ``A data-driven game-theoretic approach for behind-the-meter pv generation disaggregation,'' \emph{IEEE Transactions on Power Systems}, vol.~35, no.~4, pp. 3133--3144, 2020.

\bibitem{7784836}
F.~Ding and B.~Mather, ``On distributed pv hosting capacity estimation, sensitivity study, and improvement,'' \emph{IEEE Transactions on Sustainable Energy}, vol.~8, no.~3, pp. 1010--1020, 2017.

\bibitem{9159864}
Z.~Xuan, X.~Gao, K.~Li, F.~Wang, X.~Ge, and Y.~Hou, ``Pv-load decoupling based demand response baseline load estimation approach for residential customer with distributed pv system,'' \emph{IEEE Transactions on Industry Applications}, vol.~56, no.~6, pp. 6128--6137, 2020.

\bibitem{ERDENER2022112224}
B.~C. Erdener, C.~Feng, K.~Doubleday, A.~Florita, and B.-M. Hodge, ``A review of behind-the-meter solar forecasting,'' \emph{Renewable and Sustainable Energy Reviews}, vol. 160, p. 112224, 2022.

\bibitem{9360481}
R.~Saeedi, S.~K. Sadanandan, A.~K. Srivastava, K.~L. Davies, and A.~H. Gebremedhin, ``An adaptive machine learning framework for behind-the-meter load/pv disaggregation,'' \emph{IEEE Transactions on Industrial Informatics}, vol.~17, no.~10, pp. 7060--7069, 2021.

\bibitem{9409943}
F.~Bu, K.~Dehghanpour, Y.~Yuan, Z.~Wang, and Y.~Guo, ``Disaggregating customer-level behind-the-meter pv generation using smart meter data and solar exemplars,'' \emph{IEEE Transactions on Power Systems}, vol.~36, no.~6, pp. 5417--5427, 2021.

\bibitem{9735454}
Z.~Chen, K.~Pan, C.~S. Lai, Z.~Li, Z.~Zhao, and L.~L. Lai, ``Maximal information coefficient based residential photovoltaic power generation disaggregation,'' in \emph{2021 IEEE Sustainable Power and Energy Conference (iSPEC)}, 2021, pp. 436--441.

\bibitem{9477124}
M.~Yi and M.~Wang, ``Bayesian energy disaggregation at substations with uncertainty modeling,'' \emph{IEEE Transactions on Power Systems}, vol.~37, no.~1, pp. 764--775, 2022.

\bibitem{PAN2022118450}
K.~Pan, Z.~Chen, C.~S. Lai, C.~Xie, D.~Wang, X.~Li, Z.~Zhao, N.~Tong, and L.~L. Lai, ``An unsupervised data-driven approach for behind-the-meter photovoltaic power generation disaggregation,'' \emph{Applied Energy}, vol. 309, p. 118450, 2022.

\bibitem{9684963}
F.~Wang, X.~Ge, Z.~Dong, J.~Yan, K.~Li, F.~Xu, X.~Lu, H.~Shen, and P.~Tao, ``Joint energy disaggregation of behind-the-meter pv and battery storage: A contextually supervised source separation approach,'' \emph{IEEE Transactions on Industry Applications}, vol.~58, no.~2, pp. 1490--1501, 2022.

\bibitem{LIU2022107887}
Y.~Liu, C.~Liu, W.~Wang, S.~Gao, and X.~Huang, ``A robust non-intrusive load disaggregation method with roof-top photovoltaics,'' \emph{Electric Power Systems Research}, vol. 208, p. 107887, 2022.

\bibitem{9766031}
K.~Pan, Z.~Chen, C.~S. Lai, C.~Xie, D.~Wang, Z.~Zhao, X.~Wu, N.~Tong, L.~Lei~Lai, and N.~D. Hatziargyriou, ``A novel data-driven method for behind-the-meter solar generation disaggregation with cross-iteration refinement,'' \emph{IEEE Transactions on Smart Grid}, vol.~13, no.~5, pp. 3823--3835, 2022.

\bibitem{10125003}
A.~Stratman, T.~Hong, M.~Yi, and D.~Zhao, ``Net load forecasting with disaggregated behind-the-meter pv generation,'' \emph{IEEE Transactions on Industry Applications}, vol.~59, no.~5, pp. 5341--5351, 2023.

\bibitem{challu2023nhits}
C.~Challu, K.~G. Olivares, B.~N. Oreshkin, F.~G. Ramirez, M.~M. Canseco, and A.~Dubrawski, ``Nhits: Neural hierarchical interpolation for time series forecasting,'' in \emph{Proceedings of the AAAI conference on artificial intelligence}, vol.~37, no.~6, 2023, pp. 6989--6997.

\bibitem{vaswani2017attention}
A.~Vaswani, ``Attention is all you need,'' \emph{Advances in Neural Information Processing Systems}, 2017.

\bibitem{ratnam2017residential}
E.~L. Ratnam, S.~R. Weller, C.~M. Kellett, and A.~T. Murray, ``Residential load and rooftop pv generation: an australian distribution network dataset,'' \emph{International Journal of Sustainable Energy}, vol.~36, no.~8, pp. 787--806, 2017.

\bibitem{NREL}
N.~R.~E. Laboratory, ``Nsrdb: National solar radiation database.''

\bibitem{weinberger2009distance}
K.~Q. Weinberger and L.~K. Saul, ``Distance metric learning for large margin nearest neighbor classification.'' \emph{Journal of machine learning research}, vol.~10, no.~2, 2009.

\bibitem{breiman2001random}
L.~Breiman, ``Random forests,'' \emph{Machine learning}, vol.~45, pp. 5--32, 2001.

\bibitem{ismail2020inceptiontime}
H.~Ismail~Fawaz, B.~Lucas, G.~Forestier, C.~Pelletier, D.~F. Schmidt, J.~Weber, G.~I. Webb, L.~Idoumghar, P.-A. Muller, and F.~Petitjean, ``Inceptiontime: Finding alexnet for time series classification,'' \emph{Data Mining and Knowledge Discovery}, vol.~34, no.~6, pp. 1936--1962, 2020.

\bibitem{KARIM2019237}
F.~Karim, S.~Majumdar, H.~Darabi, and S.~Harford, ``Multivariate lstm-fcns for time series classification,'' \emph{Neural Networks}, vol. 116, pp. 237--245, 2019.

\bibitem{CHEN2021126}
W.~Chen and K.~Shi, ``Multi-scale attention convolutional neural network for time series classification,'' \emph{Neural Networks}, vol. 136, pp. 126--140, 2021.

\end{thebibliography}
\end{document}